\documentclass[twocolumn,showpacs,showkeys]{revtex4}
\usepackage{graphicx}
\usepackage{amsmath}
\usepackage{amssymb}
\usepackage{latexsym}
\usepackage{dcolumn}
\usepackage{bm}
\usepackage{url}
\usepackage{listings}

\begin{document}


\author{Aymeric Fouquier d'H\'erou\"el}
\email{afd@kth.se}
\surname{Fouquier d'H\'erou\"el}
\affiliation{Computational Biology, AlbaNova University Center, Royal Institute of Technology, Stockholm, Sweden}
\date{\today}
\title{QPS -- quadratic programming sampler,\\
a motif finder using biophysical modeling}

\begin{abstract}
 We present a Markov chain Monte Carlo algorithm for local alignments of nucleotide sequences aiming to infer putative transcription factor binding sites, referred to as the quadratic programming sampler.
 The new motif finder incorporates detailed biophysical modeling of the transcription factor binding site recognition which arises an intrinsic threshold discriminating putative binding sites from other/background sequences.
 
We validate the principal functioning of the algorithm on a sample of four promoter regions from \emph{Escherichia coli}.
 The resulting description of the motif can be readily evaluated on the whole genome to identify new putative binding sites.
\end{abstract}



\maketitle

\section{Introduction}

\begin{figure}
  \begin{center}
    \includegraphics[scale=0.25]{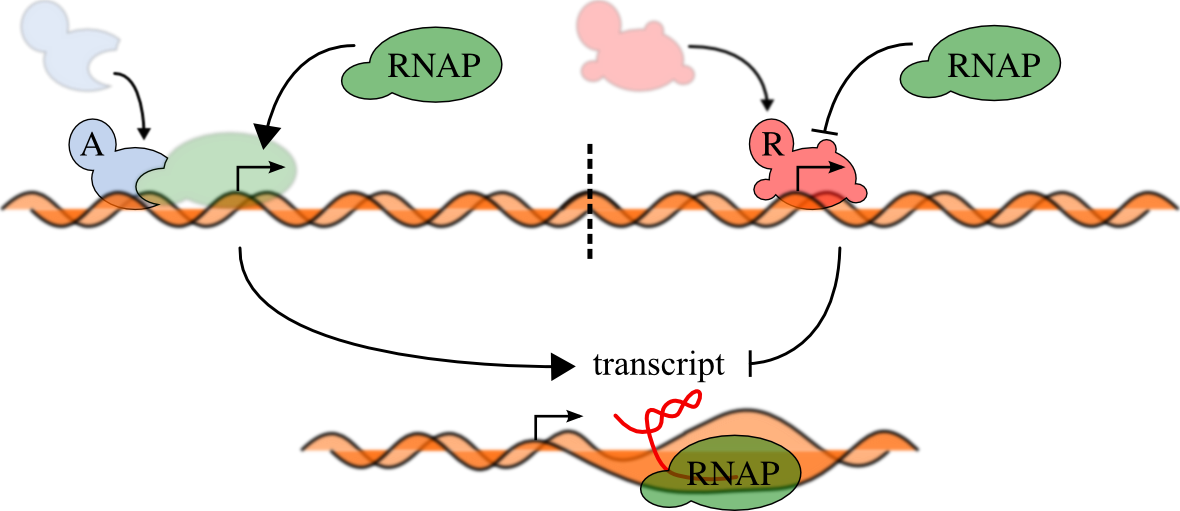}
  \end{center}
  \caption{RNA polymerase (RNAP) activity can be controlled by activating ({\bf A}) and/or repressing ({\bf R}) transcription factor proteins. Combinations of binding sites for {\bf A} and {\bf R} lead to more complex schemes of regulation.}
  \label{fig:tf_rnap}
\end{figure}

 Transcription factors (TFs) are DNA binding proteins with regulatory effects.
 They may either independently or in an interplay with other proteins activate or repress the expression of genes related to the sequence they bind to, as illustrated in Figure \ref{fig:tf_rnap}.
 In this simplified picture, they act by either facilitating or impeding the recruitment of RNA polymerase holoenzymes, protein complexes responsible for the transcription of DNA to RNA.
 Information on exact locations and sequences of TF binding sites, typically 8-15 nucleotides, not only reveals which genes may or may not be controlled by a specific TF, thus permitting the construction of networks of genetic interaction, but is also indispensable when predicting novel putative binding sites with sequence motifs defined by the known examples.
 Exact binding sequences are often still unknown as typically in \emph{Escherichia coli}, where roughly 70 of a total of 231 activating and repressing TFs have experimentally verified binding motifs \cite{regulondb,ecocyc}.
 In eukaryotes the picture is, as expected, worse: the commercial database TRANSFAC{\textregistered} contains in its most recent version 10018 entries for eukaryotic TFs, of which just 834 have reported binding motifs \cite{transfac}.

 In principle, sequences associated to coregulated or even homologous genes can be used to infer putative TF binding sites by merely aligning those sequences.
 Several methods have been proposed to perform this task efficiently, see \cite{discovering} for a summary of the most popular approaches.
 In practice, however, each of these methods has its flaws \cite{tompa,sandve}.

 The approach we present here is grounded on the representation of TFs by free energy matrices as developed in QPMEME by Djordjevic \emph{et al.} \cite{qpmeme}.
  This representation yields the contributions of specific nucleotides to the total free energy of interaction between the TF and a sequence of DNA.
 Energies are scaled in terms of the chemical potential, rendering an intrinsic binding threshold which simplifies the task of distinguishing possible binding sites from the background on genomic sequences.

\section{Method}

We proceed by reviewing a simple model of TF-DNA binding and the representation of binding site motifs before discussing the probabilistic model on which our approach is based, as well as the details of the algorithm.

\subsection{TF-DNA Binding}

The interaction between a TF and a specific sequence of DNA can be written as a pair of ordinary differential equations, describing the variation of bound and free concentrations of both reactants.
 Such a model depends on the reaction rates for binding and dissociation of TFs with DNA, symbolically stating
\begin{equation}
{\rm TF}~+~{\rm DNA}\begin{array}{c}K_{bind}\\\rightleftharpoons\\K_{diss}\end{array}{\rm TF\circ DNA}
\end{equation}
with equilibrium constants for binding and dissociation $K_{\rm bind}$ and $K_{\rm diss}$, respectively.
 In a system with many particles, the equilibrium concentrations of free TFs, specific DNA sequences and bound TF$\,\circ~$DNA complexes can be related by the Arrhenius equation
\begin{equation}
\frac{K_{\rm bind}}{K_{\rm diss}}=\frac{[{\rm TF\circ DNA}]}{[{\rm TF}][{\rm DNA}]} \sim \exp(-\beta E(S))
\end{equation}
where $\beta$ is an inverse temperature and $E(S)$ stands for the free energy of binding the TF to DNA with a specific sequence $S$.
 Accepting the probability for the TF binding the sequence $S$ to be be given by
\begin{equation}
P_{\rm b}(S)=\frac{[{\rm TF\circ DNA}]}{[{\rm TF\circ DNA}]+[{\rm DNA}]}\,,
\end{equation}
it follows
\begin{equation}
P_{\rm b}(S) = \frac{1}{1+{\rm e}^{\beta[E(S)-\mu]}}
\end{equation}
with the chemical potential $\mu$, relating abundance of the TF and affinity to its binding sites.
 The chemical potential is given by
\begin{equation}
\mu = k_{\rm B}T\log{[{\rm TF}]}+C\,,
\end{equation}
up to the additive constant $C$.

\subsection{Binding Site Motifs}

To define a binding motif from a collection of known binding sites $S_1,\dots,S_N$, of length $L$ each, the construction of a matrix $w_{\eta i}$ containing statistical weights for the occurrence of a nucleotide $\eta$ at position $i$ in the motif is usually adopted \cite{bvh}.
\begin{equation}
w_{\eta i} = \log\frac{f_{\eta i}}{p_\eta} \quad {\rm with} \quad f_{\eta i} = \frac{c_{\eta i}+1}{N+4}
\end{equation}
are constructed by counting the occurrences $c_{\eta i}$ of nucleotide $\eta$ at position $i$ in each of the binding sites and comparing the thus defined frequencies $f_{\eta i}$ to the probabilities $p_\eta$ with which to expect $\eta$ in the sequence.
 Those probabilities can be deduced from the whole genome in question, from shorter regions containing the binding sites or even just from the latter.
 $f_{\eta i}$ usually takes into account the error due to the finite amount of sequences in the collection by adding \emph{pseudocounts} \cite{bvh,karplus} to the occurrence counts.
 The weight matrix construction can then be used to evaluate the information content of a motif denoted by $w$
\begin{equation}
I_w \equiv \sum_{\eta i}w_{\eta i}f_{\eta i}\,,
\end{equation}
a measure for the dissimilarity between the motif and random sequences stemming from the probabilistic model defined by $p_\eta$.
 Further, $w_{\eta i}$ can be applied to find putative binding sites in a genome.
 Each subsequence of length $L$ is thus associated to an \emph{information score} \cite{stormo98}, describing the likelihood of that sequence to belong to the set of binding sites.
 How to chose a threshold score discriminating putative binders from non-binders, however, remains an open question in the weight matrix approach [threshold].

 A more subtle description of binding motifs by free energy matrices \cite{qpmeme} addresses the problem of finding a threshold by inverting the interpretation of a binding motif.
 Instead of describing similarities in a set of sequences, one attempts to model the requirements of a sequence to be able to bind a specific TF, now itself represented by the motif.
 The construction of such energy matrices $\varepsilon_{\eta i}$ is based upon the assumption that binding motifs represented by $\varepsilon$ should maximize the probability of recovering the set $\mathfrak M=\{M_1,M_2,\dots,M_N\}$ of known binding sites from an ensemble of random sequences, while the probability of identifying binding sites on unrelated random sequences is minimized.
  This can be performed maximizing the likelihood
\begin{equation}
\mathfrak{L}[\mathfrak{M}] = \prod_{S \in \mathfrak M} {\rm Pr}(S)P_{\varepsilon}(S) \prod_{S' \notin \mathfrak M}\left[1-{\rm Pr}(S')P_{\varepsilon}(S')\right]\,,
\end{equation}
with probabilities ${\rm Pr}(M)$ of generating a binding site sequence $M$ and binding probabilities $P_{\varepsilon}(M)$ for a TF to bind to this sequence.

 Maximizing $\mathfrak L$ can be shown to be equivalent to minimizing the variance $\sigma^2_{\varepsilon}$ of free energies resulting from the TF ($\varepsilon$) binding to random sequences \cite{qpmeme}, leading to an optimal $\varepsilon$ by solving
\begin{equation}
\arg\min_{\varepsilon} \sigma^2_{\varepsilon} \qquad {\rm subject~to} \qquad E_{\varepsilon}(M) \leq R_0\,,
\end{equation}
where $R_0$ is a threshold free energy defining binding sites.
 
 In the inference method described by Djordjevic \emph{et al.} \cite{qpmeme}, $\varepsilon$ is efficiently approximated in the low temperature limit $\beta\to\infty$ by \emph{quadratic programming}.
 The elements of the energy matrix are shifted by the mean free energy of the TF being bound to random sequences $\langle E \rangle_\varepsilon$ and rescaled by the absolute value of its chemical potential $\mu$.
 The evaluation of $\varepsilon_{\eta i}$ on a specific nucleotide sequence $M=(\alpha_1,\alpha_2,\dots,\alpha_w)$ gives the at first sight somewhat cumbersome result
\begin{equation}
R_\varepsilon(M) \equiv \sum_{i=1}^w \varepsilon_{\alpha_i i} = \frac{E_\varepsilon(S)-\langle E \rangle_\varepsilon}{\mid\mu-\langle E \rangle_\varepsilon\mid}\,,
\end{equation}
where $E_\varepsilon(M)$ is the free energy of a TF associated to a TF $\varepsilon$ binding to $M$.
 Yet this representation has a major advantage as compared to weight matrices: the chemical potential discriminates between strong and weak binding sites and since $\varepsilon_{\eta i}$ is directly inferred in terms of $\mu$, the threshold is implicitly given as $R_0 = -1$.
 All sequences with $R_\varepsilon(M) \leq R_0$ are thus presumably strong binding sites, while $R_\varepsilon(M) > R_0$ denotes weak and non-binders. More explicitly, the probability to find a specific TF bound to motif sequence $M$ is
\begin{equation}
\label{eq:pbind}
P_\varepsilon(M) = \frac{1}{1+{\rm e}^{\hat\beta[R_\varepsilon(M)+1]}}
\end{equation}
with the rescaled inverse temperature
\begin{equation}
\hat\beta = \frac{\mid\mu-\langle E \rangle_\varepsilon\mid}{k_{\rm B}T}\,.
\end{equation}

 For further details we also refer to \cite{affinity,qpmeme,selex} and references therein.
 Note, however, that $\hat\beta$ remains a free parameter as long as estimates for the average energy and the chemical potential are missing.
 Varying $\hat\beta$ does obviously not change the qualitative result in \eqref{eq:pbind} stating $P_\varepsilon>0.5$ if $R_\varepsilon>R_0$, but will lead to a sharper discrimination of binding sequences from non-binding ones.

\subsection{Probabilistic Model}
Let us introduce a first order Markov model for the genomic background with conditional probabilities for the generation of a sequence $M=(\alpha_1,\alpha_2,\dots,\alpha_w)$ written as
\begin{equation}
{\rm Pr}(\eta\mid\nu) \triangleq {\rm Pr}({\rm find} ~ \eta ~ {\rm preceded ~ by} ~ \nu)
\end{equation}
where $\eta$ and $\nu$ represent single nucleotides.
 The probability of finding $M$ among random sequences is thus given by
\begin{equation}
{\rm Pr}(M) = \prod_{i=1}^L {\rm Pr}(\alpha_i\mid\alpha_{i-1})\,,
\end{equation}
understanding the boundary condition
\begin{equation}
{\rm Pr}(\alpha_1 \mid \alpha_0) \equiv {\rm Pr}(\alpha_1)\,.
\end{equation}
Adopting a more compact notation, we introduce the \emph{passage matrices} $\mathfrak W_i(\beta)$ of probabilities for a TF to be bound to a site featuring the pair of nucleotides $\nu\eta$ at position $i$
\begin{equation}
\mathfrak W_{\nu i}^\eta(\beta) = \sum_\zeta \mathfrak \delta_{\zeta\nu}\exp(-\beta\varepsilon_{\nu i})\,{\rm Pr}(\zeta\mid\eta)\,.
\end{equation}
 Products $\prod\mathfrak W_i(\beta)$ of \emph{passage matrices} apparently yield the probabilities of coming across TF-DNA hybrids of corresponding length.
 Consequently, the generating function for a motif sequence bound by $\varepsilon$ can be written as trace of the matrix product
\begin{equation}
\mathfrak Z_\varepsilon(\beta) = {\rm tr}\left(\prod_{i=1}^L \mathfrak W_i(\beta)\right)\,,
\end{equation}
from which common statistical quantities describing binding of the TF to DNA can be derived.
 Of special interest is clearly the variance of free energies of a TF binding to random sequences, expressed as second derivative of the generating function
\begin{equation}
\sigma^2_\varepsilon = \left.\frac{\partial^2}{\partial\!\beta^2} \log\mathfrak Z_\varepsilon(\beta)\,\right|_{\beta=0}\,.
\end{equation}
Evaluating this expression, Djordjevic \emph{et al.} show how to solve for $\varepsilon$ by minimizing the variance \cite{qpmeme}.

\subsection{Monte Carlo Sampling}

 In the set of $N$ sequences $\{S_1,\dots,S_N\}$, e.g. promoter regions of co-regulated genes or upstream regions of homologous genes, we want to identify locally conserved subsequences of length $w$, supposedly sharing common TF binding sites.
 The alignment of the motif sequences $M_i$ is represented by their positions $a_i$ on the respective sequence $S_i$.
 Let us first introduce the alignment probability distribution $\hat P_\varepsilon(x)$ for a binding motif at position $x$ on sequence $S=(\alpha_1,\alpha_2,\dots,\alpha_L)$, which is constructed from the binding probabilities $P_\varepsilon$ by setting
\begin{equation}
\hat P_\varepsilon(x) = \frac{P_\varepsilon((\alpha_x,\alpha_{x+1},\dots,\alpha_{x+w}))}{\sum_{x'=1}^{L-w+1} P_\varepsilon((\alpha_{x'},\alpha_{x'+1},\dots,\alpha_{x'+w}))}\,.
\end{equation}
\begin{figure}[!ht]
  \begin{center}
    \includegraphics[scale=0.4]{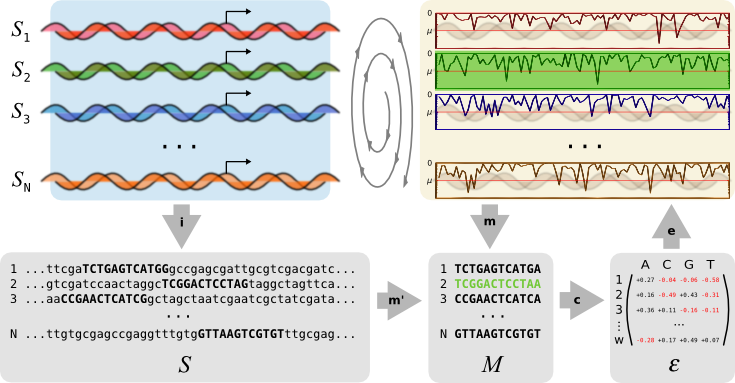}
  \end{center}
  \caption{Schematics of the QPS algorithm taking $N$ input sequences $S_1,\dots,S_N$ with initialisation ({\bf i}), motif extraction ({\bf m}), matrix computation ({\bf c}) and evaluation ({\bf e}) steps. $S_2$ is highlighted being retained for updating.}
  \label{fig:qps}
\end{figure}
 
 The inference of an optimal local alignment is accomplished by a standard Monte Carlo method following the procedure
\begin{description}
\item{({\bf i})} assign random alignment positions $a_i$, $i=1,\dots,N$
\item{({\bf m})} extract sequence motif $\mathfrak M$ of $N-1$ sequences, excluding $S_i$ where $a_i$ is to be updated
\item{({\bf c})} compute energy matrix $\varepsilon$ of the sequence motif $M$
\item{({\bf e})} evaluate $\varepsilon$ on the excluded sequence $S_i$ using $R_\varepsilon$
\item{({\bf m}')} draw new alignment position $a_i$ from the alignment probability distribution $\hat P_\varepsilon(a)$ and iterate with ({\bf c}),
\end{description}
\begin{figure}[!htb]
	\begin{center}
		\includegraphics[scale=0.35]{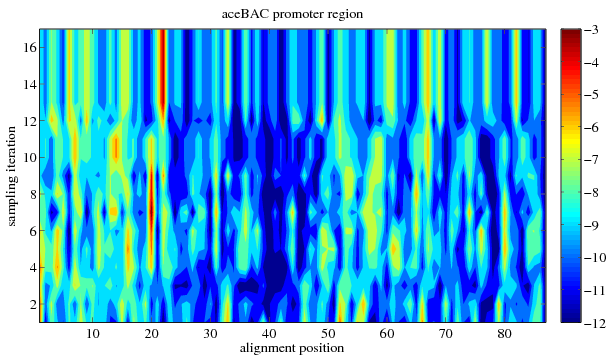}\\
		\includegraphics[scale=0.35]{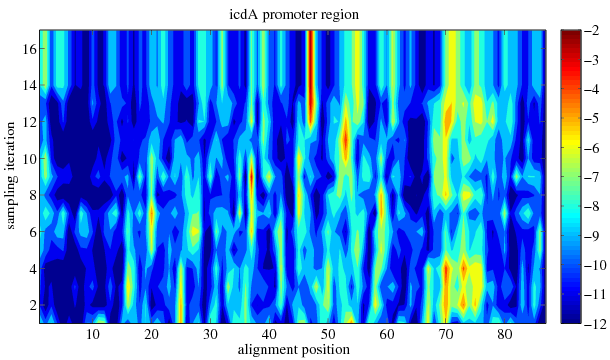}\\
		\includegraphics[scale=0.35]{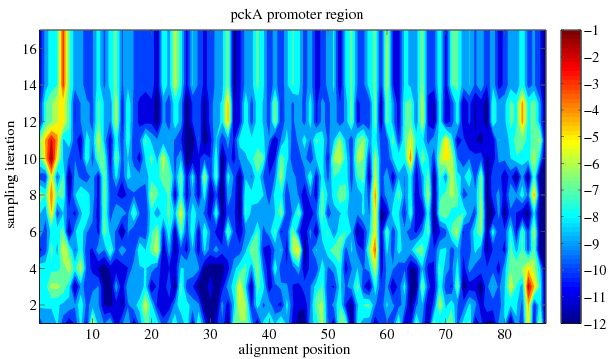}\\
		\includegraphics[scale=0.35]{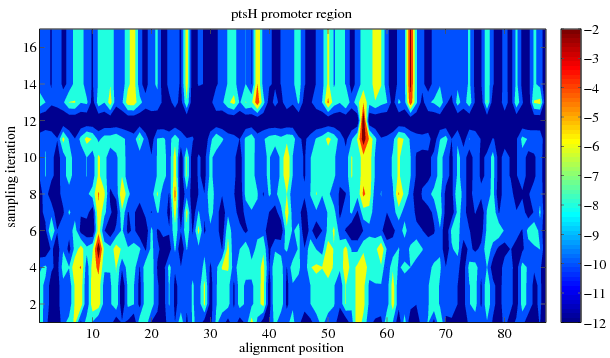}\\
	\end{center}
	\caption{Heat maps of the evolution of alignment position probability distributions on the promoter regions of different operons in \emph{Escherichia coli}. The regions contain one known TF binding site for FruR each and were aligned by QPS. At iteration 14, the distributions have reached their stationary form.}
	\label{fig:convergence}
\end{figure}

 Figure \ref{fig:qps} illustrates the procedure in which each iteration draws a new alignment on the skipped sequence.

Iteratively updating the alignments, we sample the distribution of alignment positions until reaching stationarity on all of the sequences $S_i$.
 The evolution of the distribution on biological sequences when inferring a binding motif of length 15 is shown in figure \ref{fig:convergence}.

The sampling is performed at finite temperature ($\hat\beta>0$) in equation \eqref{eq:pbind}.
 Varying the rescaled inverse temperature during sampling allows to define a simple annealing schedule with stronger discrimination of possible binding from non-binders in the actual model as $\hat\beta$ grows.

\section{Results and Discussion}

We validated the functionality of QPS on a small set of coregulated promotors in in \emph{Escherichia coli} consisting of aceBAKp, icdAp, pckAp, and ptsHp.
 Each region contains an experimentally known binding site for the fructose repressor protein FruR, which we tried to infer.
\begin{table}[!ht]
  \begin{center}
    \begin{tabular}{lcc}
      \hline
      region & $a$ & sequence\\
      \hline\hline
      aceBAK & 24 & CCTCATGCGCTTCTG\\
      icdA & 49 & GCTGAATCGCTTAAC\\
      pckA & 7  & CCCAAAGCGCCTTTT\\
      ptsH & 66 & GCTGAATCGATTTTA\\
      \hline
    \end{tabular}\\[10pt]
    $\frac{\varepsilon_{\rm FruR}}{10}=\left(
      \begin{array}{cccc}
+0.3768 & -0.2691 & -0.4809 & +0.3768\\
+0.3890 & +0.4208 & -1.1658 & +0.3596\\
+0.3890 & +0.4208 & -0.0730 & -0.7332\\
-0.0761 & -0.2755 & -0.0043 & +0.3596\\
-1.1689 & +0.4208 & +0.3921 & +0.3596\\
-0.7724 & +0.4208 & +0.3921 & -0.0369\\
+0.3890 & -0.4408 & +0.3921 & -0.3367\\
+0.3890 & +0.4208 & -1.1658 & +0.3596\\
+0.3890 & -1.1371 & +0.3921 & +0.3596\\
+0.1072 & +0.4208 & -0.8839 & +0.3596\\
+0.3890 & +0.4208 & -0.0730 & -0.7332\\
+0.3890 & +0.4208 & +0.3921 & -1.1983\\
-0.0254 & +0.4208 & -0.0043 & -0.3874\\
-0.0254 & +0.4208 & +0.3921 & -0.7839\\
+0.1072 & +0.0243 & -0.0223 & -0.1056
      \end{array}\right)
    $
  \end{center}
  \caption{QPS alignment of the four regions sharing a FruR binding site including the resulting energy matrix of the corresponding TF.}
  \label{tab:alignment}
\end{table}
 The heatmaps in figure \ref{fig:convergence} illustrate the development of the alignment probability distribution $\hat P_\varepsilon$ when iterating with QPS and one of the obtained alignments is presented in table \ref{tab:alignment}.
 We settled for initiating the algorithm with $\hat\beta=20$ and allowing for subsequent linear augmentation as the sampling proceeds,
\begin{equation}
\hat\beta_k \sim k
\end{equation}
with sampling iteration $k$.
 This corresponds approximately to an annealing schedule $\sim T^{-1}$ for the temperature.

 Exact binding sites remaining unknown for a vast number of TFs, and it is an interesting problem to try to infer a binding motif by aligning a set of sequences which are supposed to share sites for a specific TF.
 A wide range of different approaches have been developed \cite{tfbs,discovering}, greedy pattern search algorithms \cite{weeder}, context free grammar constructors \cite{mobydick}, and several statistical methods \cite{lawrence,meme}, to cite but just a small selection.
 Still it appears that no single method is capable of identifying motifs in a reliable way \cite{tompa,limitations} and more recent approaches tend to combine several algorithms \cite{harbison,discovering} to get a certain degree of cross-validation between individual methods.
 The method we present has been conceptually verified on a small sample of \emph{Escherichia coli} promoter regions and might prove useful in combination with other approaches.
 The advantage of our algorithm is that it makes direct use of a biophysical representation of the TF.
 This representation is provided as result and can be readily applied to predict yet unknown binding sites elsewhere on the genome.\\\\
 The here described algorithm has been implemented in \verb|C++| and is publicly available under the GPL on \url{http://www.csc.kth.se/~afd/qps/}.

\begin{acknowledgments}
I would like to thank Erik Aurell for useful discussions and advices on the organisation of the article.
This work was supported by the Swedish Research Council through contract number 2003-4614.
\end{acknowledgments}

\bibliographystyle{apsrev}
\bibliography{refs_qps}
\end{document}